# Jahn-Teller-driven Phase Segregation in $Mn_xCo_{3-x}O_4$ Spinel Thin Films

Miles D. Blanchet[1], Bethany E. Matthews[2], Steven R. Spurgeon[2,3], Steve M. Heald[4], Tamara Issacs-Smith[1] and Ryan B. Comes[1]

[1]Department of Physics, Auburn University, Auburn, AL 36849, USA

[2]Energy and Environment Directorate, Pacific Northwest National Laboratory, Richland, WA 99354, USA

[3]Department of Physics, University of Washington, Seattle, WA 98195, USA

[4]Advanced Photon Source, Argonne National Laboratory, Argonne, IL 60439, USA

a) Electronic mail: ryan.comes@auburn.edu

Transition metal spinel oxides comprised of Earth-abundant Mn and Co have long been explored for their use in catalytic reactions and energy storage. However, understanding of functional properties can be challenging due to differences in sample preparation and the ultimate structural properties of the materials. Epitaxial thin film synthesis provides a novel means of producing precisely-controlled materials to explore the variations reported in the literature. In this work, $Mn_xCo_{3-x}O_4$ samples from x = 0 to x = 1.28 were synthesized through molecular beam epitaxy and characterized to develop a material properties map as a function of stoichiometry. Films were characterized via *in situ* X-ray photoelectron spectroscopy, X-ray diffraction, scanning transmission electron microscopy, and polarized K-edge X-ray absorption spectroscopy. Mn cations within this range were found to be octahedrally coordinated, in line with an inverse spinel structure. Samples largely show mixed $Mn^{3+}$ and $Mn^{4+}$ character with evidence of phase segregation tendencies with increasing Mn content and increasing $Mn^{3+}$ formal charge. Phase segregation may occur due to structural incompatibility between cubic and tetragonal crystal structures associated with $Mn^{4+}$ and Jahn-Teller active $Mn^{3+}$ octahedra, respectively. Our results help to explain the reported differences across samples in these promising materials for renewable energy technologies.

## I. INTRODUCTION

Materials that exhibit oxygen reduction reactivity (ORR) are important for the development of catalytic technologies such as electrolyzers, metal-air batteries and fuel cells [1–7]. While materials containing expensive elements such as platinum dominate current catalysis research[1,3,8,9], cheaper alternatives such as transition metal oxide spinels containing manganese, cobalt, nickel, and iron have been shown to exhibit ORR levels on par with platinum-based materials[1,10–12]. This includes the spinels of the cobalt-manganese system, $CoMn_2O_4$ (CMO) and $MnCo_2O_4$ (MCO) [1,11,13–16], where previous work has demonstrated their impressive ORR properties[17–20]. While both CMO and MCO are relatively understudied in literature, an investigation of CMO thin film material properties from ideal to Co-rich cation stoichiometry was recently published[21]. The properties of MCO in literature are contentious, with different studies reporting a range of lattice parameters [22–25] and valence character [26–29]. Some studies also show phase segregation implying that MCO does not have a stable single phase [12,30]. As will be discussed, these differing results are likely due to variance in the number of $Mn^{3+}$ and $Mn^{4+}$ cations in $Mn_xCo_{3-x}O_4$ samples. Our current work here involves the material characterization of $Mn_xCo_{3-x}O_4$ films with a wide range of cation stoichiometry from ideal $Co_3O_4$ to Mn-rich MCO (x = 0 to x = 1.28).

Transition metal spinel oxides materials have chemical formula $AB_2O_4$. A-site and B-site cations generally possess a 2+ and 3+ formal charge, respectively. In rare cases, these cations can take on 4+ and 2+ charge states in what are known as "4-2" spinels[31]. A primary argument in this work is that ideal single-phase, stoichiometric MCO is a 4-2 spinel with cation valence states of $Mn^{4+}$ and $Co^{2+}$ rather than $Mn^{2+}$ and $Mn^{3+}$. Most spinels exhibit a face-centered cubic Bravais lattice structure, although some spinels take on tetragonal structures due to the Jahn-Teller (JT) effect [4,32–35]. CMO is well-known to possess a tetragonal structure due to the JT effect associated with $Mn^{3+}$ cations in octahedral coordination [1,18–21,32,33,36,37]. JT distortion does not occur in the case of $Mn^{4+}$ in octahedral coordination, however. In contrast, MCO has been determined to be a cubic spinel by many groups[38,39]. The cubic spinel unit cells can be seen in Figure 1[40]. Conventional in-plane lattice parameters for spinels with 1st row transition metals range between 8 to 9 Å [41], but the JT distortion can cause out-of-plane parameters to lie past 9 Å [21].

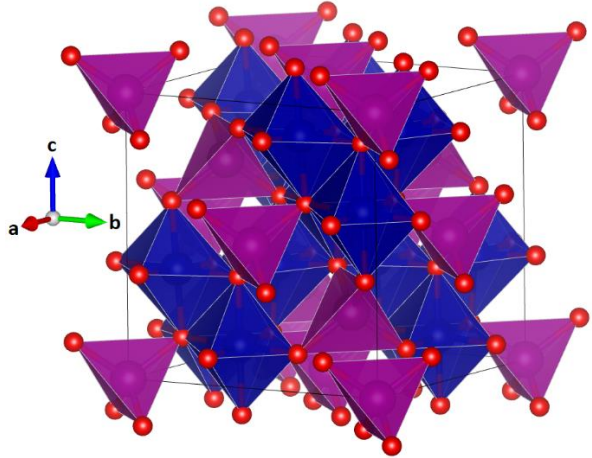

Figure 1. Conventional unit cell for a cubic structure $AB_2O_4$ spinel, with purple and blue polyhedra representing A and B cations, respectively.

A normal-type spinel is one in which all A-site cations are tetrahedrally coordinated, and all B-site cations are octahedrally coordinated. CMO is predominately normal-type in which all Co and Mn cations are tetrahedrally and octahedrally coordinated, respectively [36,37,42,43]. The tetragonal structure and high c-parameters of CMO is associated with its normal-type configuration, since it puts $Mn^{3+}$ in octahedral coordination which leads to the JT effect. $Co_3O_4$ is also known to trend towards a normal-type spinel, with $Co^{2+}$ ions in the A-site and $Co^{3+}$ in the B-site[43,44]. An inverse-type spinel is one in which all A-site cations and half of the B-site cations are octahedrally coordinated, and half of the B-site cations are tetrahedrally coordinated. MCO is predominately inverse-type, [26,45–47] with Co cations split between tetrahedral and octahedral sites. For inverse-type MCO, all Mn cations are octahedrally coordinated just as they are in CMO. This means that the material would be JT-active if Mn is in a 3+ state.

There have been numerous studies of MCO in nanocrystalline form [1,4,12,19,48,49], but very few studies involving thin films samples[46,50] and none reporting the growth of MCO using molecular beam epitaxy (MBE). However, studies involving the experimental characterization have shown mixed results in literature. Some studies report single-phase samples and others report phase segregation and mixed-valence composition [12,26–30]. These include studies reporting $Mn^{4+}$ character[26–29] and others showing only $Mn^{3+}$ [46]. This suggests that it is challenging to synthesize high-quality single-phase MCO or that a single-crystalline version of MCO may be metastable. Also, conflicting reports on material properties such as cation valence indicates that there is room for studies that investigate the basic properties of MCO.

While 2+ and 3+ are the cation valences of most spinels, single-crystalline MCO may instead possess $Mn^{4+}$ and $Co^{2+}$ states. In the case of standard $Mn^{2+}$ and $Co^{3+}$, inverse-type MCO would place $Co^{3+}$ into tetrahedral coordination which is rarely observed. This is due to the unfavorable configuration of unpaired $e_g$ orbitals associated with $Co^{3+}$ in tetrahedral coordination[51]. This leads to the idea that Co may instead take on a much more favorable 2+ valence with Mn adjusting itself to $Mn^{4+}$ for charge balancing against $O^{2-}$. Other studies of MCO showing $Mn^{4+}$ character support this idea [26–29] although without single-phase samples and reliable characterization, it is difficult to confirm. This study seeks to characterize MCO for its material properties including cation valences and investigate its propensity for phase segregation, while also providing the first demonstration of thin film synthesis by MBE.

Understanding MCO of an ideal $MnCo_2O_4$ composition is critical to its future use in energy storage technology, but studying how the material properties change with stoichiometry is also important. This includes how Co- and Mn-rich MCO behaves, but also investigating compositions that trend far towards $Co_3O_4$ and $CoMn_2O_4$. For this reason, $Mn_xCo_{3-x}O_4$ samples with a wide range of Co-Mn ratios were grown and studied from $Co_3O_4$ to $CoMn_2O_4$ with numerous samples in between. Findings from these samples act as a characterization map showing the material properties as a function of stoichiometry for the entire Co-Mn spinel system. For the purposes of this study, the MCO-region and CMO-region monikers indicate samples whose $Mn_xCo_{3-x}O_4$ stoichiometry lies below and above x = 1.5, respectively.

## II. EXPERIMENTAL

MCO films were grown on (001)-oriented MAO spinel substrates (a = 8.083 Å, MTI Corporation) using MBE. Substrates were sonicated in acetone and isopropyl alcohol for ~5 minutes each before being loaded into the MBE chamber. Mn and Co metals were deposited concurrently during growth and effusion cells were kept at constant temperature, with deposition rates calibrated using a quartz-crystal microbalance pre-growth. The sample stage was heated to a constant temperature using an infrared ceramic heating source and measured via a thermocouple on the stage, which causes an overestimation of ~50-100 ºC relative to the substrate surface temperature. Samples were grown at 500 ºC setpoints and subsequently cooled to ambient temperatures over ~30 minutes. Oxygen gas was introduced into the chamber and maintained at set flow rates during film growth and cooling which resulted in slight variation of oxygen pressures between samples, with pressures in the range of 2-3×10$^{-5}$ Torr. Significantly lower oxygen pressures are required during cooldown to prevent surface reconstruction of MCO as in CMO. A radio-frequency plasma source at 300 W power was used to significantly increase oxygen reactivity via formation of atomic O radicals.

An *in situ* reflection high-energy electron diffraction (RHEED) system was used to monitor film growth. RHEED was used as sparingly as possible, based on previous observations that the electron beam adversely affects film quality for CMO synthesis[21]. All X-ray photoelectron spectroscopy (XPS) data were collected *in vacuo*[52] using a PHI 5400 system refurbished by RBD Instruments equipped with a monochromater. The system is connected through vacuum transfer line with the MBE growth chamber, allowing sample surfaces to be measured before exposure to atmosphere. An electron neutralizer was used during measurements and spectra energy shifted to place O 1s region peaks at 530 eV.

Atomic force microscopy (AFM) topography maps were obtained using a Park XE7 AFM system in non-contact mode. Rutherford backscattering (RBS) was performed using a 6HDS-2 tandem, National Electrostatics Corporation Pelletron through helium nuclei bombardment. X-ray diffraction (XRD), reciprocal space maps (RSM), and X-ray reflectivity (XRR) data were gathered using a Rigaku SmartLab diffractometer system with a Cu Kα source with a two-bounce Ge(220) monochromator. X-ray absorption spectroscopy (XAS) fluorescence data was collected at the Advanced Photon Source at Sector-20 BM. Both in-plane (parallel) and out-of-plane (perpendicular) polarized spectra were created at both the Co and Mn K edges.

Cross-sectional scanning transmission electron microscopy (STEM) samples were prepared using a FEI Helios 600 NanoLab DualBeam $Ga^+$ Focused Ion Beam (FIB) microscope with a standard lift out procedure. STEM imaging and energy-dispersive X-ray spectroscopy (EDS) mapping was performed on a probe-corrected Thermo Fisher Themis Z microscope operating at 300 kV, with a convergence semi-angle of 25.2 mrad and an approximate collection angle range of 65–200 mrad for high-angle annular dark field (STEM-HAADF) images.

## III. RESULTS AND DISCUSSION

### A. Film Synthesis and Initial Characterization

RHEED patterns of all samples after cooling revealed them to contain spinel-structure phases, although some patterns show rough or defect-rich surfaces. Example RHEED patterns from key samples discussed in this work are shown in Figure 2. Figure 2(a) shows a high-quality $Co_3O_4$ surface, while Figures 2(b-c) show $x = 0.52$ and $x = 1.02$ samples. Samples generally showed RHEED patterns with increasing streakiness as $x$ increased and degraded in quality for $x$ greater than 1. As will be seen, these lower-quality RHEED patterns may correlate closely with phase segregation or mixed-valence character of some samples. Stoichiometry of samples was determined by RBS fitting (supplement). Film thickness was determined through fitting of XRR (supplement).

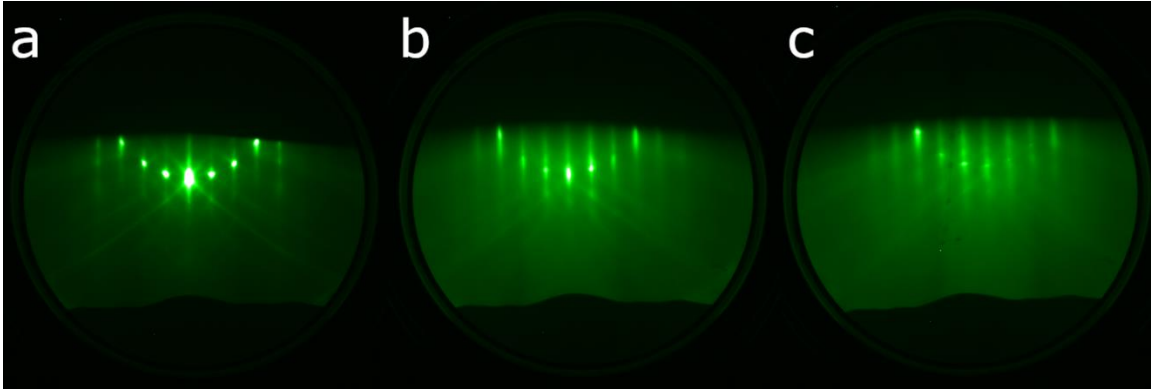

Figure 2. RHEED patterns for an a) high quality surface of $Co_3O_4$ and $Mn_xCo_{3-x}O_4$ (MCO) films with b) x = 0.52 and c) x = 1.02.

$Co_3O_4$ is on the furthest Co-rich end of the Co-Mn spinel system and is a useful starting point for a discussion on MCO's properties as a function of stoichiometry. $Co_3O_4$'s lattice parameter of 8.086 Å[53] is very close to the MAO substrates' of 8.083 Å leading to a layer-by-layer growth mode with very low lattice strain. High-quality single-crystalline samples of $Co_3O_4$ were grown and studied using a variety of characterization methods. This confirmed known properties of the ideal material but also provided a baseline for studying how the Co spinel changes with the introduction of Mn. The growth of $Co_3O_4$ thin films by MBE has been reported in literature before including synthesis using MAO substrates [54,55].

High-quality RHEED images taken through the $Co_3O_4$ growth process showed the film to appear single-crystalline with very few defects, and perfectly strained to the substrate (Figure 2(a)). X-ray diffraction of the film is consistent with a cubic FCC spinel structure, and the OOP (004) peak is convolved with the substrate indicating the small lattice mismatch between film and substrate (Figure 3(b)). XPS measurements show a Co 2p spectrum with a satellite feature indicating the mixed 2+ and 3+ character of $Co_3O_4$, which is comprised of 1/3 $Co^{2+}$ and 2/3 $Co^{3+}$ (Figure 4(a)). This satellite shape is consistent with other studies involving XPS of $Co_3O_4$ [56].

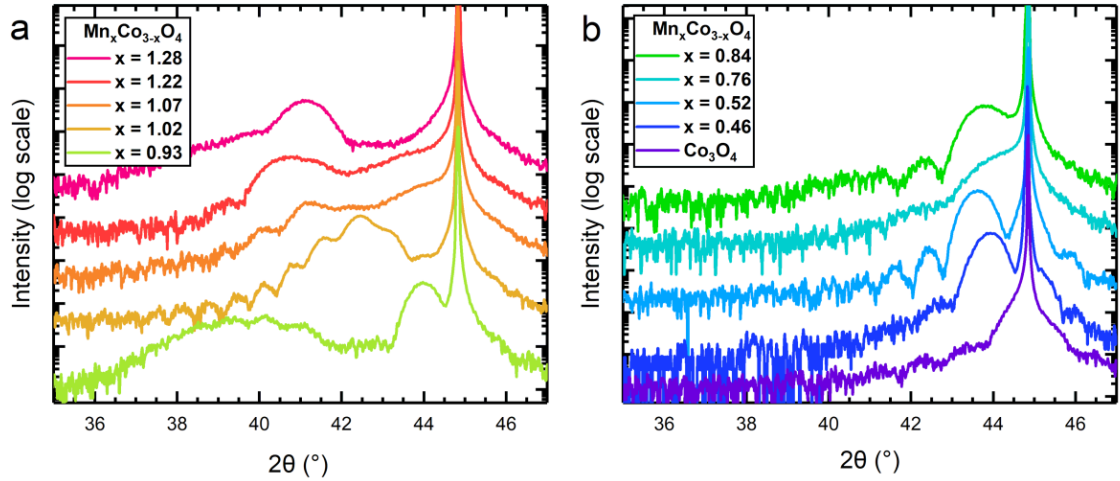

Figure 3. Out-of-plane XRD spectra of MCO-region samples that are a) phase segregated showing multiple film peaks and b) single-phase showing a single (004) spinel film peak.

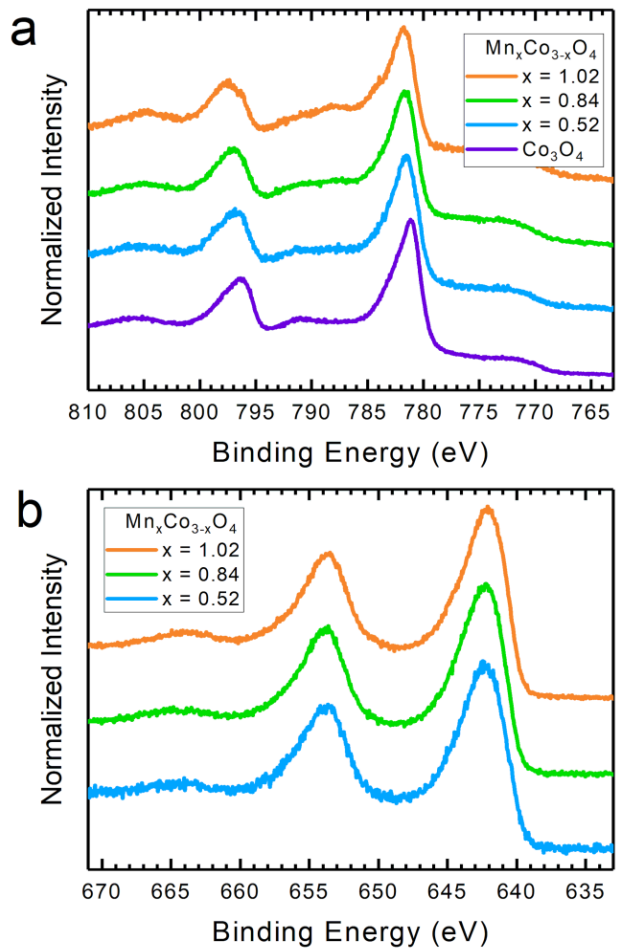

Figure 4. *In situ* XPS spectra for (a) Co 2p and (b) Mn 2p core levels for selected samples.

Starting with the characterization results of $Co_3O_4$ helps in discussing changes in material properties that comes with the introduction of Mn into the spinel system towards $MnCo_2O_4$. MCO-region ($x < 1.5$) samples tended to exhibit phase segregation behavior with increasing Mn content. Phase segregation presented itself most clearly in OOP XRD spectra in which secondary film peak signals were apparent in some samples. Location, intensity, and broadness of peaks varied greatly among phase segregated samples with some examples being shown in Figure 3(a). It is not clear if all peaks are due to the (004) spinel diffraction condition or if some reflect a material with a different crystal structure, such as rocksalt MnO or CoO. Peaks at values of $2\theta$ between 37° and 40° are consistent with tetragonal $Co_{1-y}Mn_{2+y}O_4$ films that undergo the JT distortion from a cubic spinel. As will be shown later from XAS analysis, even samples that appear to be single-phase in XRD, such as $x = 1.02$, show a tendency to display mixed Mn valence character between $Mn^{3+}$ and $Mn^{4+}$ and may have localized tetragonal distortions.

Three possible contributions can be given for the rising c-parameter. One, Mn ions generally exhibit a larger ionic radius than Co ions with the same valence state[57], leading to longer bond lengths and unit cells. Two, ionic radii of cations and thus bond lengths change depending on their valence state and coordination environment. Samples with octahedral Mn with valence character trending towards $Mn^{3+}$ (1.96 Å or 2.025 Å) rather than $Mn^{4+}$ (1.91 Å) will show greater lattice constants[57]. Three, JT distortion of $Mn^{3+}$ octahedra causes immense stretching along the out-of-plane lattice direction as seen in CMO[21]. As will be shown later through multiple analyses, Mn cations appear to occupy octahedral coordination sites in MCO-region samples. Greater Mn content means a higher percentage of $Mn^{3+}$ octahedra in samples with mixed valence character. Previous studies of ideal stoichiometry $MnCo_2O_4$ show a reported cubic structure with a lattice parameter of 8.09 Å to 8.29 Å [22–24], but the sample in this study closest to ideal MCO ($x = 1.02$) shows an out-of-plane lattice parameter of 8.50(1) Å, suggesting some degree of JT activity despite the apparent single phase based on XRD.

*In situ* XPS of MCO-region samples indicate that all Mn is either in a $Mn^{3+}$ or $Mn^{4+}$ state, with no $Mn^{2+}$ found in any sample (Figure 3(b)). The satellite feature characteristic of $Mn^{2+}$, which would be found between the $2p_{1/2}$ and $2p_{3/2}$ peaks, is not seen in the Mn 2p spectra [58,59]. Mn 2p spectral shapes of $Mn^{3+}$ and $Mn^{4+}$ appear identical and thus even qualitative determination of Mn valence from this region was not attempted. XPS also shows a general trend from mixed $Co^{2+}$ and $Co^{3+}$ character towards predominately $Co^{2+}$ character with increasing Mn content. This is seen through the changing shape of the satellite feature of the Co 2p XPS region, which is located between the $2p_{1/2}$ and $2p_{3/2}$ peaks (Figure 4(b)). The changes show progressively less indication of $Co^{3+}$ in higher-Mn samples, with some samples even depicting ideal $Co^{2+}$ satellite features [21].

## *B. Electron Microscopy*

To further probe phase segregation in the samples, STEM and EDS measurements were performed on samples $x = 0.76$ and $x = 1.02$. These results are shown in Figures 5 and 6 respectively. STEM high-angle annular dark-field (HAADF) images (Figure 5(a)), in which contrast is proportional to atomic number $Z^{\sim 1.7}$, confirm the overall uniformity and sharpness of the film, while STEM bright-field (BF) (Figure 5(b)) images show no significant strain or other defects, such as phase separation. STEM-EDS compositional analysis (Figure 5(c)) for $x = 0.76$ also confirms a homogeneous sample with uniform structure and no evidence of Co and Mn segregation in the regions examined.

In contrast, the STEM-EDS map for $x = 1.02$ (Figure 6(c)) shows that Co and Mn are distributed heterogeneously within the film, with some Mn cations having migrated towards the sample surface. However, no clear secondary phases are observed, with the exception of a surface reconstruction in the top few unit cells, which can likely be attributed to atmospheric exposure, as indicated by STEM-HAADF and STEM-BF images (Figs. 6.a-b). The cation segregation in this sample that has close-to-ideal $MnCo_2O_4$ stoichiometry leads to the question of whether a stable single-phase version of the material exists, or whether it will seek to form multiple phases as seen in samples of other studies [12,30].

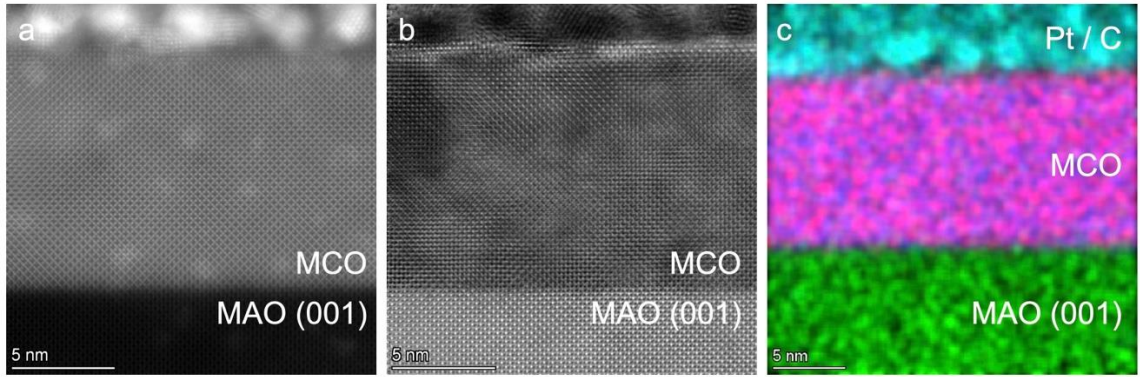

Figure 5. Scanning transmission electron microscopy images of $x = 0.76$ sample. (a) Drift-corrected STEM high-angle annular dark-field (HAADF) image; (b) bright-field (BF) image; (c) composite STEM-EDS map of sample showing Co $K$ (blue), Mn $K$ (red), Mg $K$ (green), and Pt $L$ (cyan) peaks. The mottled contrast in the HAADF is from Pt redeposition during TEM preparation. Images taken along the MAO [100] zone-axis.

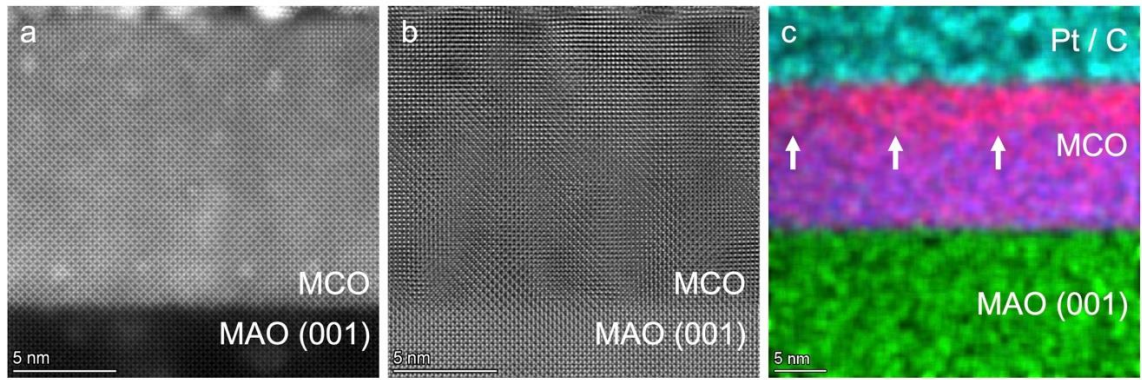

Figure 6. Scanning transmission electron microscopy images of $x = 1.02$ sample. (a) Drift-corrected STEM high-angle annular dark-field (HAADF) image; (b) bright-field (BF) image; (c) composite EDS map of sample showing Co $K$ (blue), Mn $K$ (red), Mg $K$ (green), and Pt $L$ (cyan) peaks. The arrows in c mark the relative increase in Mn content near the film surface. The mottled contrast in the HAADF is from Pt redeposition during TEM preparation. Images taken along the MAO [100] zone-axis.

## C. X-ray Absorption Spectroscopy

XAS was performed on $Co_3O_4$ along with other MCO-region samples over a range of film stoichiometries. Incident X-rays were linearly polarized allowing for the acquisition of data sensitive to either the in-plane and out-of-plane lattice directions only. A large pre-edge peak is seen in spectra of both polarizations near the Co K-edge in Figure 7(a), indicating tetrahedral coordination of Co cations in the sample [60]. As will be seen, this pre-edge peak appears in Co XANES spectra for all MCO-region samples, indicating that tetrahedral sites remain occupied by Co and additional Mn occupies octahedral sites. This is also consistent with the presumed inverse-type spinel structure of MCO.

As was the case in CMO, analysis of the EXAFS region allowed for determination of cation-nearest oxygen bond lengths in $Co_3O_4$ and other MCO-region samples (supplement). However, Co cations are both tetrahedrally and octahedrally coordinated which means that spectra contain information from both lengths and fitting must account for both contributions. This can be done by introducing scattering paths for both octahedral and tetrahedral coordination into the fits and adjusting the weighting values in the fits to account for the multiple sites.

One possible combination of lengths for tetrahedral $Co^{2+}$ and octahedral $Co^{3+}$, respectively, is 1.91(1) Å and 1.90(1) Å from the in-plane spectrum and 1.92(1) Å and 1.91(1) Å from the out-of-plane spectrum. Despite the fact that these values for tetrahedrally coordinated $Co^{2+}$ are lower than the value of 1.96 Å predicted through theoretical crystal radii [57], fitting error was low (supplement) and similar bond lengths have been reported in other EXAFS analyses of $Co_3O_4$ [61,62]. Crystal strain is unrelated to this shrinking in tetrahedral bond length since the lattice mismatch between substrate and film is minor. As will be seen, similar bond lengths were determined for other Co spectra of MCO-region samples indicating that values did not vary greatly with the addition of Mn.

Assuming additional Mn takes on octahedral coordination as indicated in multiple analyses of this study, this lowering of Co valence could correspond with charge transfer from Mn to Co. In this case, $Mn^{4+}$ occupation of octahedral sites would produce a corresponding $Co^{2+}$ for charge balancing, thereby leading to idealized $MnCo_2O_4$ valences of $Mn^{4+}$ and $Co^{2+}$. As will be shown, many of the MCO-region samples of this study show mixed valence character between $Mn^{3+}$ and $Mn^{4+}$. This indicates that both octahedral site substitutions may take place, with JT-active $Mn^{3+}$ leading to higher c-lattice parameters. The phase segregation behavior of MCO-region seems closely tied to mixed-valence behavior, suggesting that $Mn^{4+}$ is present in single-phase MCO while $Mn^{3+}$ cations lead to phase segregation.

In-plane and out-of-plane polarization XAS was performed on MCO-region samples with a range of stoichiometry from $Co_3O_4$ to Mn-rich MCO. Out-of-plane Mn spectra show significant variation with changing stoichiometry through the appearance of a pre-edge feature at ~6550 eV of the K-edge (Figure 7(e)). This pre-edge feature intensity is similar in appearance to the pre-edge features shown in out-of-plane Mn spectra of epitaxial tetragonal CMO samples[21]. This indicates that changing properties between samples, including phase segregation tendency and multi-valence character, is driven by Mn but not Co. This also suggests that the changing material properties are rooted in causes that are anisotropic along the out-of-plane direction.

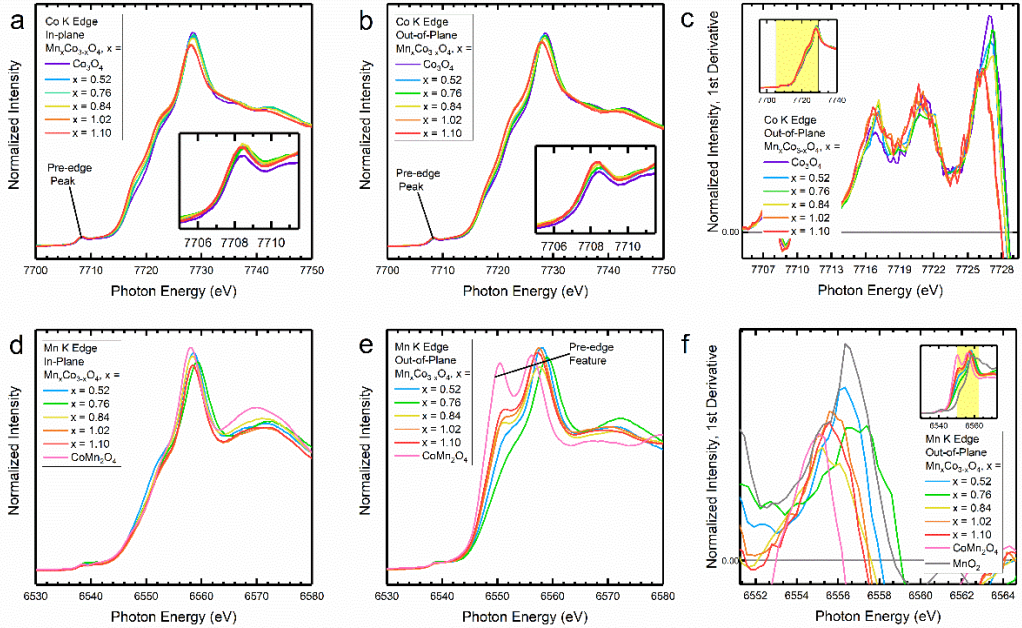

Figure 7. Co K edge XANES data for (a) in-plane and (b) out-of-plane polarizations, with (c) 1st derivative of out-of-plane data. Mn K edge XANES data for (d) in-plane and (e) out-of-plane polarizations with (f) 1st derivative of out-of-plane data.

$Mn^{3+}$ character most likely explains the appearance of the pre-edge feature. Based on in-plane polarization measurements, most of these MCO-region samples exhibit some mixed Mn valence character between 3+ and 4+, while CMO samples is comprised solely of $Mn^{3+}$. Considering the relatively large c-lattice parameters of these MCO-region samples and the tetragonal crystal structure of CMO, this pre-edge feature may be a general indication of JT distortion of $Mn^{3+}$ octahedra and of a tetragonal crystal structure. Also, the existence of these pre-edge features further supports the idea that all Mn cations in MCO-region samples take on octahedral coordination. One $Mn_xCo_{3-x}O_4$ sample, x = 0.76, shows no pre-edge feature intensity in its out-of-plane Mn spectrum indicating that this sample is comprised predominately of $Mn^{4+}$ as seen in other analyses as well. While this single-phase, single-valence sample is Co-rich compared to $MnCo_2O_4$ stoichiometry, it may be an indication that ideal MCO does have a stable single phase with $Mn^{4+}$ and $Co^{2+}$ valence states.

Co-region XANES for all samples and both polarizations show high-intensity pre-edge peaks at the base of the K-edge (insets of Figure 7(a-b)). The intensities of the peaks are consistent between samples and because pre-edge intensity is an indication of tetrahedral coordination for Co[63], all tetrahedral sites appear to be fully occupied by Co regardless of sample stoichiometry. This follows for an inverse-type MCO spinel in which all Co would be found in tetrahedral sites and further supports the idea that Mn takes on octahedral coordination.

Linear combination fitting of both in-plane and out-of-plane Co region spectra shows a trend in valence character from mixed $Co^{2+}$ and $Co^{3+}$ towards $Co^{2+}$ with increasing Mn stoichiometry (see Supplemental Table S2). Fitting was performed by using $Co_3O_4$ (2.67+ Co valence) and CMO (2+ Co valence) spectra as fitting standards. This trend in Co valence confirms what was also seen in comparing Co 2p XPS spectra. Linear combination fitting of in-plane Mn shows no clear trend with varying stoichiometry and indicates the mixed-valence character of the MCO-region samples of this study that is consistent with differing degrees of phase segregation between samples. Fitting was performed using CMO (3+ Mn valence) and $MnO_2$ (4+ Mn valence) as standards and results confirm that most samples exhibit mixed $Mn^{3+}$ and $Mn^{4+}$ character. While linear combination fitting can show general trends in cation valences between samples, exact valence determination is not possible since fitted and reference spectra do not share identical structures.

Analysis of Co-region XAS derivative plots for both polarizations also shows trends in Co valence from mixed 2+ and 3+ towards $Co^{2+}$ with increasing Mn stoichiometry. MCO-region samples show cations with multiple coordinations and mixed-valence character, making EXAFS fitting a challenge. Not only do theoretical ionic radii depend on cation coordination and valence state[57], JT distortion drives large changes in bond lengths with the introduction of $Mn^{3+}$ in octahedral sites. This is especially problematic in fitting out-of-plane Mn spectra for samples that show both $Mn^{3+}$ and $Mn^{4+}$. All this leads to multiple bond lengths in the material for a given cation and must be accounted for in fitting spectra by introducing additional scattering paths into the fitting algorithm. Determining multiple bond lengths from one spectrum is difficult when the contribution of each length in a sample is not known.

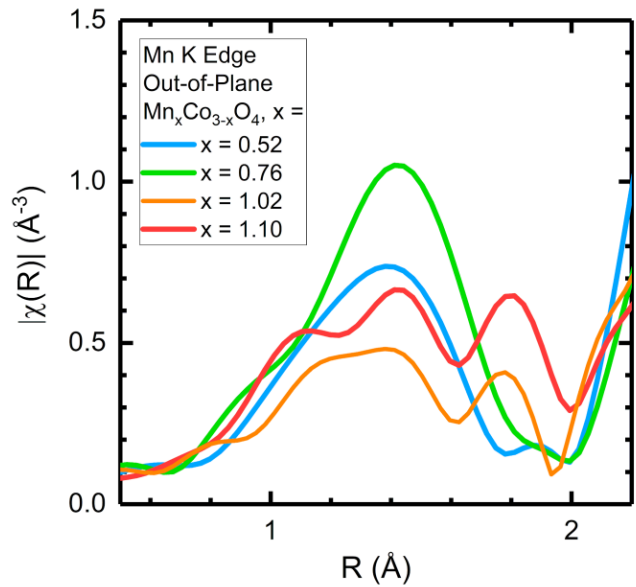

Figure 8. Mn K edge EXAFS data for out-of-plane polarization for selected stoichiometries.

The $x = 0.76$ sample does not show multiple bond lengths in its out-of-plane Mn spectrum and can be fit in a straightforward manner. This sample shows in-plane and out-of-plane Mn-O bond length values of 1.90 Å and 1.91 Å, respectively, both of which correspond closely to theoretical bond lengths for $Mn^{4+}$ [57]. This indicates, as other analyses of this study do, that this single-phase sample exhibits primarily $Mn^{4+}$ character and no $Mn^{3+}$. While its stoichiometry is Co-rich, this MCO-region sample shows no mixed valence character and may exemplify the closest to the chemical ideal of $MnCo_2O_4$, namely a Mn valence state of 4+.

All analyses suggest that Mn in the MCO-region samples of this study are octahedrally coordinated but also show mixed Mn valence character between $Mn^{3+}$ and $Mn^{4+}$. While other studies of MCO report mixed valence character as well [26–29], an ideal spinel should generally show single valence states for each cation meaning that mixed-valence character is a reflection of defects in the material. OOP XRD c-lattice parameters steadily increase with the addition of Mn indicating the action of JT distorted $Mn^{3+}$ octahedra. Co-region XANES pre-edge peaks show consistent high intensity indicating that tetrahedral sites are occupied by Co cations leaving Mn to occupy octahedra. The appearance of a pre-edge feature in out-of-plane Mn-region XANES spectra indicates $Mn^{3+}$ octahedra in MCO-region samples, since CMO also shows this feature and contains only $Mn^{3+}$ octahedra. Transformed EXAFS spectra for out-of-plane Mn spectra show multiple nearest-neighbor oxygen bond lengths which most likely comes from the simultaneous occupation of octahedral sites with both $Mn^{3+}$ and $Mn^{4+}$ cations. This octahedral coordination of Mn also confirms the inverse-type spinel nature of MCO and stoichiometry-varied samples from $Co_3O_4$ to Mn-rich MCO. While $Mn^{3+}$ may not be the ideal Mn valence for single-phase MCO, the fact that samples show this character allows the octahedral coordination of Mn to be observed in this study, primarily through the act of JT distortion.

## D. Discussion

From the findings of this study, it is clear that the $MnCo_2O_4$ phase stability is highly dependent on processing conditions and prone to phase segregation to a miscibility gap between $Co_3O_4$ and $CoMn_2O_4$ type phases. The sample closest to this stoichiometry is the $Mn_xCo_{3-x}O_4$ sample $x = 1.02$, which shows evidence of cation segregation and mixed-valence character between $Mn^{3+}$ and $Mn^{4+}$. However, a key sample in this study from the single-phase MCO-region material with $x = 0.76$ shows solely $Mn^{4+}$ character with no $Mn^{3+}$. In this sample, out-of-plane polarization Mn K edge XANES shows no indication of a pre-edge feature associated with $Mn^{3+}$ [21]. EXAFS fitting shows near-identical Mn-O bond lengths for both in-plane and out-of-plane polarization spectra, the values of which correspond closely to $Mn^{4+}$ octahedra. First derivative XAS also suggests the sample shows predominately $Mn^{4+}$ character. As discussed in the supplemental information, the x = 0.76 sample was grown at the high end of the oxygen pressure range for this study (~$3\times10^{-5}$ Torr) may explain the greater concentration of $Mn^{4+}$.

OOP XRD also shows that the c-lattice parameter of sample $x = 0.76$ is lower than those of other MCO-region samples. Lower lattice parameters have been reported previously for ideal MCO [22–24] with values trending as low as 8.09 Å. This suggests that the sample $x = 0.76$ is closer in lattice parameter to true, single-phase MCO than the other MCO-region samples of this study which show greater lattice parameters. It was discussed earlier how these other MCO-region samples show higher c-lattice parameters due to the effects of JT distortion and polyhedral stretching of octahedral $Mn^{3+}$. Thus, $Mn^{4+}$ seems to be the characteristic valence of ideal MCO while $Mn^{3+}$ represents valence state defects. These defects then act to drive the material into larger lattice parameters through JT distortion, and potential phase segregation in the case simultaneous $Mn^{3+}$ and $Mn^{4+}$ character.

While sample x = 0.76 is Co-rich, its single-phase, single-valence and smaller lattice parameter suggests that ideal $MnCo_2O_4$ does have a stable single-crystalline phase with a Mn valence of $Mn^{4+}$. Additionally, XPS and XAS show a trend in Co valence from mixed 2+ and 3+ towards $Co^{2+}$ with increased Mn stoichiometry for MCO-region samples, indicating that charge transfer from Mn to Co on the octahedral sites occurs. This would mean that studies showing mixed-valence character or phase segregation [12,26–30] involve should be viewed with this in mind.

A transition from ideal $MnCo_2O_4$ ($Mn^{4+}$ and $Co^{2+}$) towards $CoMn_2O_4$ ($Co^{2+}$ and $Mn^{3+}$) via the addition of Mn would see the substitution $Mn^{3+}$ cations into the spinel structure. While MCO possesses a cubic crystal structure with non-JT-active $Mn^{4+}$ octahedra, CMO exhibits a tetragonal structure due to its distorted $Mn^{3+}$ octahedra. Due to the incompatibility between cubic and tetragonal c-lattice parameters, the transition in stoichiometry from MCO towards CMO and replacement of $Mn^{4+}$ with $Mn^{3+}$ in octahedral sites produces stoichiometries that likely do not have a stable single phase. Instead, the coexistence of $Mn^{4+}$ and $Mn^{3+}$ may lead to regions of cubic and tetragonal structure that would favor phase segregation over a homogenous structure due to a miscibility gap in the Mn-Co spinel phase diagram. Limited work has been reported on the Co-Mn spinel phase diagram, but Golikov et al have previously observed such a miscibility gap in bulk ceramics[64]. Their work reported that the phase diagram is highly dependent on the preparation and thermal treatment process, which is consistent with our observations. The thin films that we have synthesized here are also qualitatively consistent with observations by Yang et al of phase segregation in Co-Mn spinel nanoparticles, where they observed formation of a nearly stoichiometric $Co_3O_4$ core and $Mn_3O_4$ shell[12]. We suggest that there is still a large parameter space to explore in this rich phase diagram for these technologically relevant materials.

## IV. SUMMARY AND CONCLUSIONS

This study is the first to synthesize and study MCO-region samples using MBE, and one of the only studies of MCO oriented towards thin films. The samples studied include ideal stoichiometry $MnCo_2O_4$ and $Mn_xCo_{3-x}O_4$ with x-values ranging from 0 to 1.28. The samples show increasing c-lattice parameters and phase segregation tendencies with additional Mn content. Co valence character also trends towards Co2+ from mixed 2+ and 3+ with increasing Mn. While most samples suggest mixed Mn valence between 3+ and 4+, $Mn^{3+}$ character likely represents valence state defects in MCO-region samples with $Mn^{4+}$ being the proper formal valence. This means the single-phase valence states of stoichiometric $MnCo_2O_4$ are likely $Mn^{4+}$ and $Co^{2+}$. $Mn^{3+}$ character in MCO-region samples is likely due to insufficient oxygen reactivity during growth. While $Mn^{3+}$ is not expected for phase-pure in $MnCo_2O_4$, oxygen-deficiency leads to the JT distortion of $Mn^{3+}$ octahedra and leads to phase segregation. It is this JT distortion of $Mn^{3+}$ octahedra that contributes to the large increases in c-lattice parameters of MCO-region samples. Mixed Mn valence between 3+ and 4+ also likely induces phase segregation in MCO-region samples, where $Mn^{4+}$ and $Mn^{3+}$ octahedra are associated with incompatible cubic and tetragonal crystal structures, respectively.

## ACKNOWLEDGMENTS

M.D.B. gratefully acknowledges support from the Alabama EPSCOR Graduate Research Scholars Program. M.D.B. and R.B.C. gratefully acknowledge support from the National Science Foundation (NSF) through NSF-DMR-1809847. B.E.M. and S.R.S. were supported by the Chemical Dynamics Initiative / Investment, under the Laboratory Directed Research and Development (LDRD) Program at Pacific Northwest National Laboratory (PNNL). PNNL is a multi-program national laboratory operated for the U.S. Department of Energy (DOE) by Battelle Memorial Institute under Contract No. DE-AC05-76RL01830. A portion of this research was performed at the W. R. Wiley Environmental Molecular Sciences Laboratory, a DOE User Facility sponsored by the Office of Biological and Environmental Research and located at PNNL. This research used resources of the Advanced Photon Source, an Office of Science User Facility operated for the U.S. Department of Energy (DOE) Office of Science by Argonne National Laboratory, and was supported by the U.S. DOE under Contract No. DE-AC02-06CH11357.

# Jahn-Teller-driven Phase Segregation in $Mn_xCo_{3-x}O_4$ Spinel Thin Films

Miles D. Blanchet[1], Bethany E. Matthews[2], Steven R. Spurgeon[2,3], Steve M. Heald[4], Tamara Issacs-Smith[1] and Ryan B. Comes[1]

[1] Department of Physics, Auburn University, Auburn, AL 36849, USA

[2] Energy and Environment Directorate, Pacific Northwest National Laboratory, Richland, WA 99354, USA

[3] Department of Physics, University of Washington, Seattle, WA 98195, USA

[4] Advanced Photon Source, Argonne National Laboratory, Argonne, IL 60439, USA

a) Electronic mail: ryan.comes@auburn.edu

## SUPPLEMENTAL INFORMATION

## I. ATOMIC FORCE MICROSCOPY

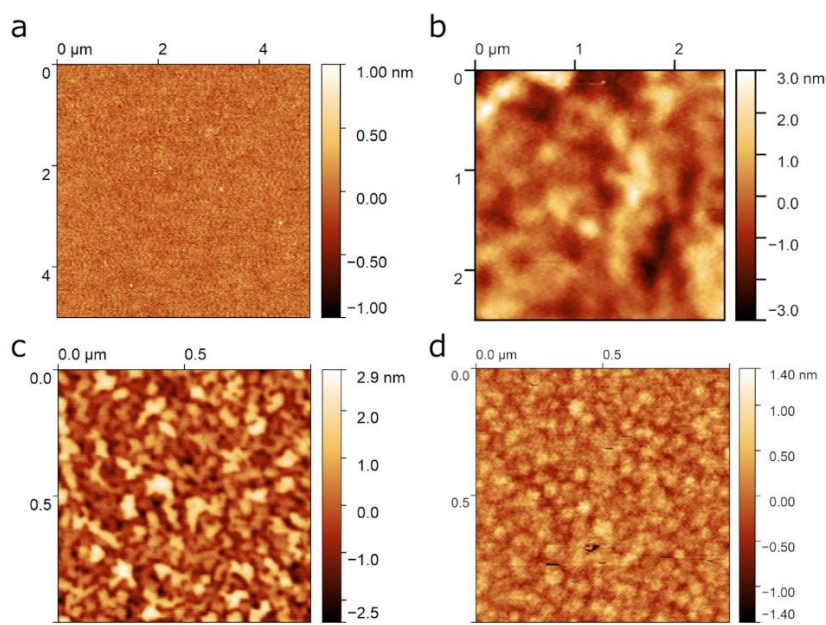

Figure S1: Atomic force microscopy topography maps of (a) x = 1.02; (b) x = 0.76; (c) x = 0.46; (d) x = 1.28.

## II. X-RAY DIFFRACTION

Notable in Table 1 is the $Mn_xCo_{3-x}O_4$ sample $x = 0.76$ which is a single-phase MCO-region sample with an OOP XRD film peak convoled with the substrate's peak. The significantly lower lattice parameter indicates that there are very few JT-active Mn octahedra and thus the sample exhibits primarily $Mn^{2+}$ or $Mn^{4+}$ character as opposed to $Mn^{3+}$. As will be shown later, multiple analyses show that the Mn valence for this sample appears to be predominately $Mn^{4+}$. It will also be argued that this single-phase sample supports the idea that $MnCo_2O_4$ does have a stable single phase that is comprised of $Mn^{4+}$ and $Co^{2+}$ as opposed to $Mn^{2+}$ and $Co^{3+}$ as a typical spinel would have. Also clear from Figure S2 are the greater c-lattice parameters of CMO-region samples which stems from the large amount of JT-distortion taking place due to Mn solely occupying octahedra as $Mn^{3+}$ in the spinel structure.

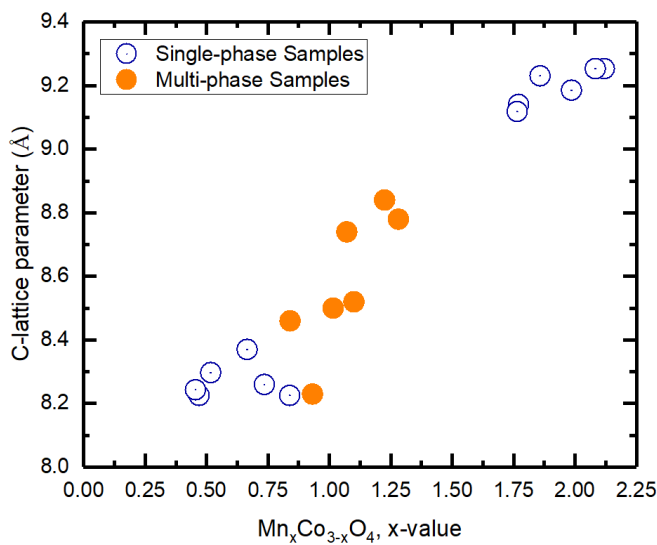

Figure S2: c-axis lattice parameters for samples of varying stoichiometry throughout the study.

## III. EXTENDED X-RAY ABSORPTION FINE STRUCTURE (EXAFS) ANALYSIS

Table S1: Fitting ratios of Mn and Co valence for in-plane XANES measurements of samples with varying stoichiometry.

| Sample | In-plane Co spectra | | In-plane Mn spectra | |
|---|---|---|---|---|
| $Mn_xCo_{3-x}O_4$, x = | % Comp. of $Co_3O_4$ ($Co^{2+}$ & $Co^{3+}$) | % Comp. of CMO ($Co^{2+}$) | % Comp. of $MnO_2$ ($Mn^{4+}$) | % Comp. of CMO ($Mn^{3+}$) |
| CMO Sample | 0 | 1 | 1 | 0 |
| 1.10 | 0.64 | 0.36 | 0.87 | 0.14 |
| 1.02 | 0.62 | 0.38 | 0.84 | 0.16 |
| 0.84 | 0.77 | 0.23 | 0.93 | 0.07 |
| 0.76 | 0.88 | 0.12 | 0.69 | 0.31 |
| 0.52 | 0.85 | 0.15 | 0.98 | 0.02 |
| $Co_3O_4$ Sample | 1 | 0 | - | - |
| $MnO_2$ Standard | - | - | 0 | 1 |

Table S2: Analysis of out-of-plane XANES data for Co and Mn.

| Sample | Out-of-plane Cobalt Spectra | | | Out-of-plane Manganese Spectra | |
|---|---|---|---|---|---|
| | Maximum Intensity Value | | Ratio | Peak Maximum Position (eV) | |
| $Mn_xCo_{3-x}O_4$, x = | 2+ Peak | 3+ Peak | 2+/3+ | Actual Position | Relative to CMO |
| CMO Sample | - | - | - | 6555.16 | 0.00 |
| 1.10 | 0.132 | 0.174 | 0.76 | 6555.58 | 0.42 |
| 1.02 | 0.127 | 0.169 | 0.75 | 6555.58 | 0.42 |
| 0.84 | 0.140 | 0.187 | 0.75 | 6555.24 | 0.08 |
| 0.76 | 0.135 | 0.214 | 0.63 | 6557.39 | 2.23 |
| 0.52 | 0.128 | 0.202 | 0.63 | 6556.34 | 1.18 |
| $Co_3O_4$ Sample | 0.111 | 0.229 | 0.49 | - | - |
| $MnO_2$ Standard | - | - | - | 6556.34 | 1.18 |

Table S3: EXAFS fitting data for bond lengths of each ionic coordination.

| Sample | Co in-plane (Å) | | Co out-of-plane (Å) | | Mn in-plane (Å) | | Mn out-of-plane (Å) | |
|---|---|---|---|---|---|---|---|---|
| $Mn_xCo_{3-x}O_4$, x = | Tetra. Coord. | Octa. Coord. | Tetra. Coord. | Octa. Coord. | Tetra. Coord. | Octa. Coord. | Tetra. Coord. | Octa. Coord. |
| 2.08 | 1.98 | - | 1.96 | - | - | 1.92 | - | 2.23 |
| 1.99 | 1.97 | - | 1.96 | - | - | 1.91 | - | 2.22 |
| 1.86 | 1.97 | - | 1.96 | - | - | 1.91 | - | 2.23 |
| 1.10 | 1.92 | 1.91 | 1.90 | 1.94 | - | 1.89 | - | Multiple |
| 1.02 | 1.92 | 1.91 | 1.94 | 1.92 | - | 1.89 | - | Multiple |
| 0.84 | 1.91 | 1.90 | 1.91 | 1.91 | - | 1.90 | - | Multiple |
| 0.76 | 1.92 | 1.90 | 1.92 | 1.90 | - | 1.90 | - | 1.91 |
| 0.52 | 1.91 | 1.90 | 1.92 | 1.90 | - | 1.90 | - | Multiple |
| $Co_3O_4$ | 1.91 | 1.90 | 1.92 | 1.91 | - | - | - | - |

All fits assume that Mn in the lattice is octahedrally coordinated so that all models assume fully tetrahedral coordination of cobalt and an appropriate number of Mn octahedra according to stoichiometry. This follows from the analysis results discussed previously that show octahedral coordination of Mn and from the nature of an inverse-type MCO spinel. Also, out-of-plane Mn spectra of most MCO-region samples show indications of multiple bond lengths due to the shape of their nearest-neighbor oxygen scattering path peaks. Figure 5.9 shows a spectrum showing a single Mn-O bond length along with two spectra showing multiple Mn-O bond lengths.

This trend is seen by comparing each derivative plot's two $Co^{2+}$ and $Co^{3+}$ peak intensity values between samples, which are found at ~7717 eV and ~7727 eV, respectively (Figure 7 in main text). Table S2 shows each peak's maximum intensity value and associated ratio between them for a given spectrum, and it is seen that ratios increase in favor of higher $Co^{2+}$ intensity with additional Mn content in a sample. While XAS derivative plots of Co-regions show two distinct peaks for $Co^{2+}$ and $Co^{3+}$, analysis of Mn-region plots is challenging since both $Mn^{3+}$ and

$Mn^{4+}$ constitute one single peak (at ~6558 eV) that shifts in energy position depending on valence state (Figure 7(f) in main text). Table S2 shows the energy position of each sample's Mn peak as well as the position relative to a CMO sample's peak position, for all in-plane spectra. CMO represents complete $Mn^{3+}$ character while peaks that trend towards higher binding energy represent increasing $Mn^{4+}$ character towards the position of a $MnO_2$ standard's peak representing $Mn^{4+}$ character. As in linear combination fitting of spectra, there are no clear trends in Mn valence with sample stoichiometry, however the peak position of $Mn_xCo_{3-x}O_4$ sample $x = 0.76$ lies at higher binding energy than the peak position of $Mn^{4+}$, indicating that this sample is predominately comprised of $Mn^{4+}$ cations as opposed to $Mn^{3+}$.

The nearest neighbor oxygen peaks are found within the range of about 1 to 2 Å. The general shape of a single-length scattering path is a single peak with a shoulder on its left side, while a peak that appears widened or misshapen appears in the case of multiple scattering paths. The peak for $x = 0.34$ depicts a single bond length while the peak for $x = 0.21$ is mishappen and contains signal from multiple bond lengths. $x = 0.58$ displays two peaks within the 1 to 2 Å radial range and is the clearest indication of multiple bond lengths in MCO-region samples. Multiple lengths appearing in out-of-plane Mn spectra would come about in the case of mixed $Mn^{3+}$ and $Mn^{4+}$ character since $Mn^{3+}$ octahedra will distort to longer bond lengths due to JT distortion alongside undistorted $Mn^{3+}$ octahedra. Thus, the existence of multiple bond lengths in out-of-plane Mn spectra shows the octahedral coordination of Mn cations in MCO-region samples.

Similar to how fitting was performed for $Co_3O_4$, two Co-O bond lengths were assumed to exist in MCO-region samples- one for tetrahedral coordination and one for octahedral. This follows from the idea that additional Mn occupies octahedral coordination in the spinel structure thereby allowing a dual-coordination cobalt structure to persist as it does in $Co_3O_4$. Co-O bond lengths determined for MCO-region samples are shown in Table 5.5 along with those determined for CMO. Bond lengths of MCO-region samples did not significantly deviate from those determined in $Co_3O_4$ despite the addition of Mn, particularly in the case of in-plane cobalt spectra. This reflects the similar shapes of Co-region XAS spectra between samples. The fact that cobalt bond lengths are relative close between samples indicates that MCO-region samples maintain a similar cobalt polyhedra fine structure to that of $Co_3O_4$. This leaves manganese cations to drive the majority of material property changes as a function of stoichiometry. The reason for an increased bond length of 1.94 Å in $x = 1.02$ and $x = 1.10$ is not known but these higher-Mn stoichiometry samples show phase segregation while the lower Mn samples are single-phase. Cobalt bond lengths in CMO samples are higher than those of MCO-region samples (ranging from 1.96 Å to 1.98 Å), and a length of 1.94 Å could follow from a stoichiometry transition from $MnCo_2O_4$ towards $CoMn_2O_4$ in which Co-O lengths increase.

Fitting for Mn-O bond lengths was performed using models that placed Mn cations in octahedral coordination only, and results are shown in Table 5.5. Bond lengths could be determined for all in-plane Mn spectra, but most out-of-plane spectra could not be fit due to spectra showing multiple bond lengths. Mixed valence character between $Mn^{3+}$ and $Mn^{4+}$ in samples is a likely contributor to multiple bond lengths through JT distortion stretching $Mn^{3+}$ octahedra but not $Mn^{4+}$, although phase segregation causing other structural changes in the sample lattices may also contribute. Fitting cobalt spectra for multiple Co-O bond lengths was achievable because the ratio of tetrahedral cobalt to octahedral cobalt was known based on sample stoichiometry, and scattering paths could be weighted properly. However, there was no way to determine the ratio of $Mn^{3+}$ to $Mn^{4+}$ cations using methods within the scope of this study and thus scattering paths would never be weighted properly.

Mn in-plane spectra were able to be fit in a straight-forward manner and showed consistent bond lengths between the values of 1.89 Å and 1.90 Å for all samples regardless of stoichiometry or phase behavior. This is interesting considering the vast differences in out-of-plane spectra in which multiple bond lengths are present due to samples possessing both $Mn^{3+}$ and $Mn^{4+}$ character. This means that both types of octahedra have relatively close in-plane bond lengths regardless of their out-of-plane lengths, although this may be due in part to lattice strain. The theoretical bond lengths based on ionic radii for $Mn^{3+}$ and $Mn^{4+}$ octahedra are 1.91 Å and either 1.96 Å (low spin) or 2.025 Å (high spin), respectively [58]. Therefore, these experimental bond lengths determined in EXAFS fitting make sense in the case of $Mn^{4+}$ octahedra, but not in the case of $Mn^{3+}$ octahedra. However, JT distortion acts to distort $Mn^{3+}$ octahedra and lead to bond lengths that are longer out-of-plane and shorter in-plane. Bond lengths associated with distorted $Mn^{3+}$ octahedra were already determined from EXAFS fitting of CMO in which fits of in-plane Mn spectra gave Mn-O lengths between 1.91 Å and 1.92 Å. This explains why in-plane spectra of MCO-region samples with mixed $Mn^{3+}$ and $Mn^{4+}$ character can be fit with a single bond length. This also follows from the fact that all in-plane Mn XAS spectra have similar shapes regardless of stoichiometry or valence character.

## IV. EFFECTS OF OXYGEN GROWTH CONDITIONS

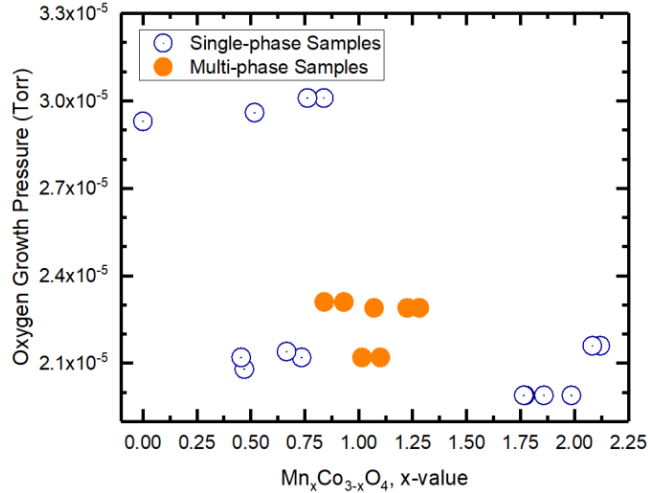

Figure S3: Oxygen growth pressure for samples examined in this study. All samples were grown with oxygen plasma active.

It was discussed earlier that phase segregation trends with increasing Mn-content in MCO-region samples. As phase segregation has been observed in the material before [12,30], this supports the suggestion that a stable single-phase version of ideal $MnCo_2O_4$ may be difficult to synthesize or does not exist entirely. However, a possible contributing factor to phase segregation in these samples are varying oxygen conditions during sample growth. Figure S3 shows oxygen pressures during growth as a function of stoichiometry, with samples' phase behavior also indicated. It can be seen that phase segregation trends with increasing Mn content, however segregation also trends with oxygen pressure used during growth. All multi-phase samples were grown at oxygen pressures below $2.5 \times 10^{-5}$ Torr while only single-phase samples were grown at oxygen pressures $\sim 3.0 \times 10^{-5}$ Torr. This indicates that phase segregation may have taken place due to insufficient oxygen reactivity during sample growth. If this is true, the mechanism may be in the formation of mixed-valence samples with both $Mn^{3+}$ and $Mn^{4+}$ in octahedral sites. Mixed-valence character was seen in most samples grown below $2.5 \times 10^{-5}$ Torr oxygen pressure, and insufficient oxygen reactivity would act to reduce cation valences due to the lower number of of $O^{2-}$ atoms in the material. $Mn^{3+}$ and $Mn^{4+}$ octahedra show significantly different out-of-plane bond lengths due to JT distortion effects, and these differently shaped polyhedra lead to opposing tetragonal and cubic crystal structures, respectively. If multiple Mn valences form during growth, separate regions comprised predominately $Mn^{3+}$ and $Mn^{4+}$ may concurrently form and naturally cause phase segregation due to physical incompatibilities between tetragonal and cubic structures.